\documentclass[aps,pra, amsmath, showpacs, preprintnumbers,superscriptaddress,twocolumn,sort&compress,floatfix, amssymb]{revtex4}
\pdfoutput=1
\usepackage{graphicx}
\usepackage{dcolumn}
\usepackage{bm}
\usepackage{color}
\usepackage{mathptmx, textcomp}
\usepackage[latin1]{inputenc}
\usepackage{braket}

\usepackage{multirow}

\begin{document}

\author{Michael Gr\"obner}\affiliation{Institut f\"ur Experimentalphysik und Zentrum f\"ur Quantenphysik, Universit\"at Innsbruck, 6020 Innsbruck, Austria}
\author{Philipp Weinmann}\affiliation{Institut f\"ur Quantenoptik und Quanteninformation (IQOQI), \"Osterreichische Akademie der Wissenschaften, 6020 Innsbruck, Austria}
\author{Emil Kirilov}\affiliation{Institut f\"ur Experimentalphysik und Zentrum f\"ur Quantenphysik, Universit\"at Innsbruck, 6020 Innsbruck, Austria}
\author{Hanns-Christoph N\"agerl}\affiliation{Institut f\"ur Experimentalphysik und Zentrum f\"ur Quantenphysik, Universit\"at Innsbruck, 6020 Innsbruck, Austria}
\author{Paul S. Julienne}\affiliation{Joint Quantum Institute, University of Maryland and National Institute for Standards and Technology, College Park, Maryland 20742, United States of America}
\author{C. Ruth Le Sueur} \affiliation{Joint Quantum Centre (JQC)
Durham-Newcastle, Department of Chemistry, Durham University, South Road,
Durham, DH1 3LE, United Kingdom }
\author{Jeremy M. Hutson}
\affiliation{Joint Quantum Centre (JQC) Durham-Newcastle, Department of
Chemistry, Durham University, South Road, Durham, DH1 3LE, United Kingdom }

\title{Observation of interspecies Feshbach resonances in an ultracold
$^{39}$K-$^{133}$Cs mixture \\ and refinement of interaction potentials}

\pacs{34.50.Cx, 34.20.Cf, 67.60.Bc}

\begin{abstract}
We observe interspecies Feshbach resonances due to s-wave bound
states in ultracold $^{39}$K-$^{133}$Cs scattering for three different spin
mixtures. The resonances are observed as joint atom loss and heating of the K
sample. We perform least-squares fits to obtain improved K-Cs interaction
potentials that reproduce the observed resonances, and carry out
coupled-channel calculations to characterize the scattering and bound-state
properties for $^{39}$K-Cs, $^{40}$K-Cs and $^{41}$K-Cs. Our results open up
the possibilities of tuning interactions in K-Cs atomic mixtures and of
producing ultracold KCs molecules.
\end{abstract}

\maketitle

\section{Introduction}

The possibility of controlling collisional interactions in ultracold atomic
samples to very high precision through Feshbach resonances \cite{Chin2010fri}
is the foundation of many different cold atom experiments. Control of
interactions by Feshbach tuning has enabled experiments on tunable quantum
gases \cite{Cornish2000,Donley2001doc, Weber2003}, the creation of ultracold
Feshbach molecules \cite{Donley2002,Regal2003cum,Herbig2003poa}, the formation
of Bose-Einstein condensates (BECs) of molecules
\cite{Jochim2003bec,greiner2003} and the observation of the BEC-BCS crossover
\cite{Regal2004,Bartenstein2004,Zwierlein2004}, few-body and Efimov physics
\cite{Kraemer2006efe}, polaron physics \cite{Schirotzek2009,koschorreck2012},
and novel states, phase transitions, and dynamics in one-dimensional gases
\cite{Haller2009roa,Haller2010pqp,Meinert2016boi}. In these experiments,
intraspecies interactions were tuned, in some cases between different spin
states of the same species. In recent years, interest has turned to mixtures of
quantum gases and the tuning of interspecies interactions. This interest is
motivated by the study of exotic phases such as supersolids
\cite{Titvinidze2008}, the heteronuclear Efimov scenario for a three-body
system \cite{Wacker2016,Ulmanis2016,Barontini2009}, boson-mediated superfluids
\cite{Efremov2002}, quantum phases that involve composite fermions
\cite{Lewenstein2004}, mixtures under simultaneous superfluidity
\cite{Ferrier-Barbut2014}, and the possibility of forming samples of ultracold
polar ground-state molecules \cite{Ni2008}. In particular, the electric dipole
moment of heteronuclear molecules gives rise to anisotropic, long-range
dipole-dipole interactions that contrast with the isotropic, short-range
interaction in atomic experiments \cite{Baranov2012,Lahaye2009}. Combining
long-range interactions with optical lattice potentials allows the study of
exotic quantum phases such as pair superfluids and the implementation of
quantum simulation and quantum information processing
\cite{Capogrosso2010,Micheli2007,Trefzger2011,Trefzger2009}.

Experimentally, systems of ultracold ground-state molecules are produced in a
two-step procedure: First, atoms in nearly quantum-degenerate atomic mixtures
are magneto-associated using a Feshbach resonance to form weakly bound
molecules. Second, these molecules are optically transferred into the
rovibrational ground state by stimulated Raman adiabatic passage (STIRAP)
\cite{Bergmann1998,Danzl2008qgo,Ni2008}. This procedure, which requires
precise knowledge of the inter- and intraspecies scattering properties, has
recently led to the production of ultracold and dense samples of heteronuclear
molecules such as fermionic KRb \cite{Ni2008} and NaK \cite{Park2015} and
bosonic RbCs \cite{Takekoshi2014,Molony2014,Reichsollner2016} and NaRb
\cite{Guo2016} in their rovibrational ground states. The present
paper is aimed towards the goal of producing ultracold KCs molecules by similar
methods. Ground-state KCs molecules are of particular interest because of
their large electric dipole moment (1.92~D) \cite{Aymar2005} and their
stability under two-body molecular collisions \cite{Zuchowski2010}, which makes
$^{40}$KCs the only chemically stable fermionic alkali-metal dimer apart from
Na$^{40}$K. Additionally, the two available bosonic isotopes $^{39,41}$K
increase the flexibility in mixing and dimer association with Cs.

In most magneto-association experiments so far, molecules were produced in three-dimensional (3D)
bulk atom mixtures \cite{Ni2008,Takekoshi2014,Molony2014,Park2015,Guo2016}.
Only a comparatively small fraction of atoms could be converted to
heteronuclear dimers, because in bulk samples the process is
limited by atomic three-body recombination and vibrational relaxation in
atom-molecule and molecule-molecule collisions. Such losses can
be suppressed if the two atomic samples are overlapped in an optical lattice,
creating either a Bose (Fermi) Mott insulator (band insulator) \cite{Moses2015} or
a Bose-Bose double-species Mott insulator \cite{Reichsollner2016}. In both
cases reported so far, a Feshbach resonance was exploited in two different
ways. First, it was used to null the interspecies interaction at the
zero-crossing of the resonance to achieve efficient sample mixing.
Subsequently, the resonance was used to form the molecules from atom pairs.
Lattice filling fractions of 30\% and above have been achieved. Since we aim at
a similar strategy for KCs, precise knowledge of the Feshbach resonance
positions and widths is crucial.

The individual two-body interaction properties of $^{39}$K \cite{Chiara2007}
and Cs \cite{Leo2000,Berninger2013} are well understood. This has allowed the
production of Bose-Einstein condensates for each species separately
\cite{Weber2003,Roati2007,Landini2012,Salomon2014} and for both species in
the same apparatus \cite{Groebner2016}. The singlet and triplet interaction
potentials for KCs have been determined from extensive electronic spectroscopy
by Ferber {\em et al.}\ \cite{Ferber:2009, Ferber2013}; in the present work we
designate the potentials of Ref.\ \cite{Ferber2013} the F2013 potentials. Patel
{\em et al.}\ \cite{Patel2014} carried out coupled-channel calculations on
the F2013 potentials to obtain the positions and widths of Feshbach resonances
for all three K isotopes, including both s-wave and d-wave bound states.
However, no experiments have yet been carried out to test these predictions. In
this article, we report the observation of Feshbach resonances in an ultracold
$^{39}$K-$^{133}$Cs mixture. We prepare the samples in different spin states
and search for loss features as we scan the homogeneous magnetic field in the
range from 0 to 650 G. The observed resonances in the lowest spin state are
observed at magnetic fields about 20~G higher than predicted in 
Refs.~\cite{Ferber2013,Patel2014}. We therefore use coupled-channel
calculations to assign the resonances and to fit improved interaction
potentials, which we designate G2017. We then use the new potentials to make
improved predictions of resonance positions and widths for all three
isotopologs of KCs.

\section{Experiment}

Techniques for the preparation of ultracold Cs \cite{Weber2003} and $^{39}$K
\cite{Landini2012,Groebner2016} are well established, but mixing the two species
is not straightforward. In particular, mixing the samples in the regime of
quantum degeneracy is quite involved and we are pursuing a strategy
similar to that demonstrated for $^{87}$RbCs in Ref.~\cite{Reichsollner2016}.
For the present goal of detecting interspecies Feshbach resonances, however, it
is sufficient to mix very cold thermal samples, and even that poses some
challenges. The different steps in laser cooling lead to a disparity in the
sample temperatures ($\geq 5~\mu$K for $^{39}$K and $\leq 1~\mu$K for Cs) and
densities. The negative background scattering length of $^{39}$K implies the 
existence of a Ramsauer-Townsend minimum in the elastic cross section 
at an energy around $400$~$\mu$K$\times k_{\rm B}$ \cite{Landini2012}, where the
scattering phase shift passes through zero and the contribution from higher
partial waves is still small. This minimum in the elastic cross section, the 
large losses when $^{39}$K atoms overlap with the Cs
magneto-optical trap (MOT), and the strong heating when Cs is loaded into a
deep dipole trap make a sequential cooling scheme necessary. We achieve this
with a translatable and transformable trap. Specifically, we first load the
$^{39}$K sample into a very tight optical trap, subsequently translate this
sample vertically to allow for Cs loading and cooling, and finally bring the
two species together in a relaxed trap with enlarged waist that is suitable for both species.

\begin{figure}
	\begin{center}{
			\includegraphics[width=1\columnwidth]{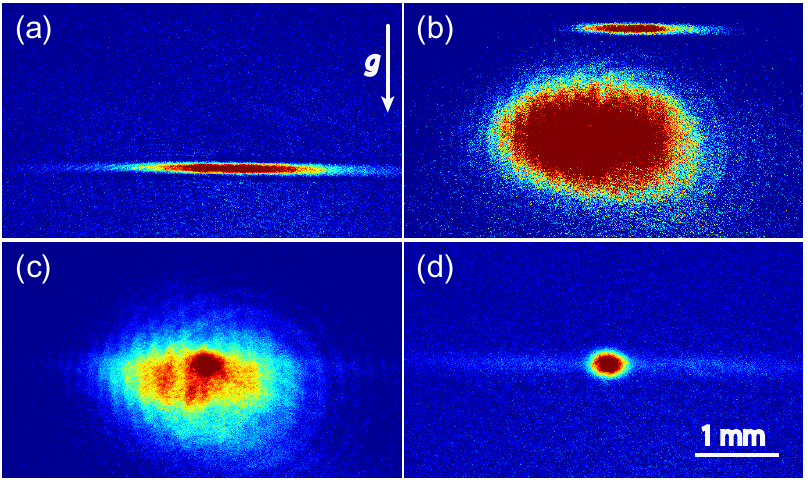}}
\caption{\label{FIG1}Experimental sequence to combine $^{39}$K with $^{133}$Cs
as shown in absorption images. (a) In situ image of the K sample after dipole
trap loading. (b) Vertically displaced K cloud (upper cloud) to avoid
collisional loss with Cs atoms (lower cloud) during Cs laser cooling. (c)
Merged samples at the end of the Cs MOT stage. The K sample shows up as the
dark red disk at the center of the image. (d) Typical K signal in the crossed
trap for a K-Cs mixture after merging. The images in (b) and (c) are overlapped
absorption images taken at the respective imaging wavelengths.}
	\end{center}
\end{figure}

The experimental sequence starts with the preparation of an ultracold K sample
as described in Ref.~\cite{Groebner2016}. In short, after standard laser
cooling and spin polarizing on the D$_2$ and D$_1$ lines, we load up to $5
\times 10^8$ atoms in the $\ket{\mathrm{K:c}} \equiv \ket{f=1,m_f=-1 }$
state into a magnetic quadrupole trap. 
Hyperfine sublevels of each atom are indicated 
by alphabetic labels a, b, c, etc., in order of increasing energy.
To overcome the Ramsauer-Townsend scattering minimum, we superimpose a dipole trap beam with $1/e^2$-waist of
26~$\mu$m at 1064~nm and an initial power of 15~W, and simultaneously increase
the quadrupole field within 5~s from 32 to 75~G/cm (along the coil axis).
Although this increases the temperature, the higher density ensures efficient
loading of the tight tweezer trap. The quadrupole field is shut off and the
magnetic offset field $B$ is then ramped to $42.5$~G. At this field the
scattering length for atoms in state $\ket{\mathrm{K:c}}$ is around 100\ $a_0$ and we
can perform efficient forced evaporative cooling. For this, the power of the
single-beam dipole trap is decreased exponentially in 1.5~s to 150~mW. During
the first 600~ms of this ramp we move the optical trap 1.2~mm upwards [see
Figs.~\ref{FIG1}(a) to \ref{FIG1}(b)]. The vertical transport is achieved by moving a lens and
a mirror that are mounted on a motorized translation stage and works without
any observable atom loss or heating. We note that magnetization effects related
to our stainless steel vacuum chamber require a magnetic polarization stage
after the quadrupole trap. Polarization is achieved by pulsing $B$ several
times up to $1000$~G for 100~ms. Without this procedure, laser cooling of Cs,
as performed subsequently, is not possible without adjustments in the magnetic
field.

At this point, the magnetic trap center is free and we can start loading the Cs
MOT. For this, we turn the quadrupole field on again (7.5~G/cm along the coil
axis). During the first 100~ms we linearly increase the K trap power to 300~mW
and turn on a 15-W dipole trap beam with a waist of 250 $\mu$m at
$\sim$1070~nm, crossing the center of the Cs MOT. After 5~s of Cs MOT loading,
and before increasing the quadrupole field to 20~G/cm to compress the Cs
sample, we superimpose the two clouds [see Fig.~\ref{FIG1}(c)]. This is done by
moving the K trap 0.79~mm downwards in 160~ms. At the same time we dynamically
increase the waist of the K trap from 26 to 63~$\mu$m by shrinking the
aperture of an iris with a servomotor and increase the power to 1.2~W. After
the compression stage the Cs sample is further cooled and spin-polarized by
three-dimensional degenerate Raman-sideband cooling (dRSC)
\cite{Treutlein2001,Kraemer2004opo}. The temperature after dRSC is below 1
$\mu$K when we release the atoms into free space. Here, however, we cool the
atoms into a crossed-dipole trap. When we do so, we measure temperatures of
about 7 $\mu$K and observe some significant atom loss. We attribute the
temperature increase and atom loss largely to the mismatch of the Cs cloud size
after dRSC to the trapping volume of the crossed-dipole trap and possibly to
ac-Stark shifts due to the dipole trap that compromise the performance of dRSC.
Also, atoms that are cooled away from the center of the crossed-dipole trap
convert potential energy into kinetic energy after extinction of the dRSC
lattice beams. In any case, after a hold time of 80~ms at 35.5~G in the
crossed-dipole trap, the K and Cs clouds [see Fig.~\ref{FIG1}(d)] are each found to be in
thermal equilibrium, but at different temperatures. They see trap depths of
about $U_{\mathrm{K}} /k_\mathrm{B}=20$~$\mu$K and
$U_{\mathrm{Cs}}/k_\mathrm{B}=39$~$\mu$K. With around $1 \times 10^5$ K atoms
and $1 \times 10^5$ Cs atoms in the trap, we measure temperatures of $T_{
\mathrm{K}} = 3$~$\mu$K and $T_{\mathrm{Cs}} = 7$~$\mu$K. We measure trap
frequencies of $\omega_{\mathrm{K}} /2\pi=(374,84,383)$ Hz and
$\omega_{\mathrm{Cs}} /2\pi=(281,63,288)$ Hz and deduce atomic peak
densities of $n_{ \mathrm{K}} = 1.2 \times 10^{12}$~cm$^{-3}$ and
$n_{\mathrm{Cs}} = 9 \times 10^{11}$~cm$^{-3}$. At this stage, the K atoms are
fully spin polarized in the third-lowest energy state $\ket{\mathrm{K:c}}$. The
Cs atoms are 80\% polarized in the $\ket{\mathrm{Cs:a}} \equiv \ket{f=3,m_f=3}$
state, with the rest of the Cs sample mainly populating the
$\ket{\mathrm{Cs:b}} \equiv \ket{f=3,m_f=2}$ state.

\begin{figure}
	\begin{center}{
			\includegraphics[width=1\columnwidth]{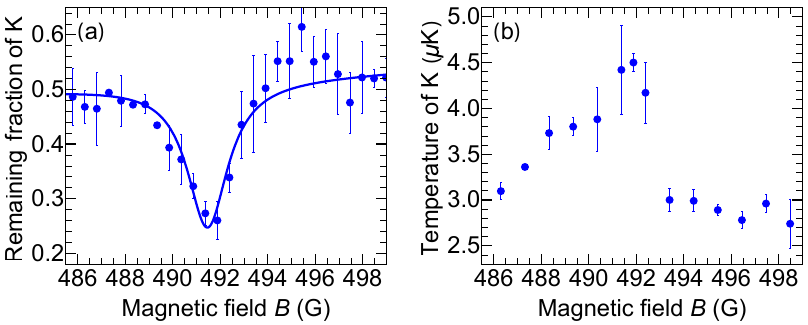}}
		\caption{\label{FIG2}Typical Feshbach resonance signatures in a
			$\ket{\mathrm{K:c}}$-$\ket{\mathrm{Cs:a}}$ mixture. (a) Normalized atom number
			and (b) temperature for a K sample mixed with Cs after a 900-ms hold time. (a)
			The remaining fraction of K atoms normalized to a sample without Cs at constant
			trap depth after a hold time of 100~ms. Each point is an average of at
			least two measurements and the solid line is a Lorentzian fit to the results. The
			temperatures in (b) are deduced from five time-of-flight images with different
			expansion times and the error bars represent statistical errors from the
			temperature fits. Larger data samples were not possible due to drifts of
			the vacuum chamber magnetization as discussed in the text.
		}
	\end{center}
\end{figure}

The dipole trap now allows us to prepare K-Cs mixtures in any desired hyperfine
state combination. Here, to provide sufficient input to theoretical modeling,
we are mainly interested in combining $\ket{\mathrm{Cs:a}}$ with
$\ket{\mathrm{K:a}}\equiv \ket{f=1,m_f=1}$, $\ket{\mathrm{K:b}}\equiv
\ket{f=1,m_f=0}$, and $\ket{\mathrm{K:c}}$. We fully spin-polarize the Cs
sample into the $\ket{\mathrm{Cs:a}}$ state by using a microwave pulse in
combination with resonant light to clean out the $\ket{\mathrm{Cs:b}}$
population. During the pulse, which lasts 6~ms and is resonant with the
transition from $\ket{\mathrm{Cs:b}}$ to $\ket{f=4}$, we sweep $B$ from 1 to
1.05~G to address magnetic field inhomogeneities, and apply laser light on the
$6^2$S$_{1/2}$ $\ket{f=4} \rightarrow$ $6^2$P$_{3/2}$$\ket{f'=5}$ transition.

%%%%%%%%%%%%%%%%%%%%%%%%%%%%%
\begin{table}[t]
	\caption[]{Overview of interspecies Feshbach resonances for mixtures
		$\ket{\mathrm{K:a}}$-$\ket{\mathrm{Cs:a}}$,
		$\ket{\mathrm{K:b}}$-$\ket{\mathrm{Cs:a}}$, and
		$\ket{\mathrm{K:c}}$-$\ket{\mathrm{Cs:a}}$. Experimentally we deduce the
		positions $B_{\rm res}$ and FWHM $\delta$ by fitting Lorentzian functions to the loss
		features. The uncertainties are the statistical errors from the Lorentzian
		fits. Note that the Lorentzian width $\delta$ is not the same physical quantity
		as the theoretical width $\Delta$. We note that drifts of the chamber
		magnetization result in a systematic error of up to $0.3$~G for $B_{\rm res}$.
	}
	\label{TABLE1}
	\begin{ruledtabular}
		\begin{center}
			\begin{tabular}{cllcrc}
				\multicolumn{3}{c} {Experiment}                   &  &  \multicolumn{2}{c} {Theory (F2013 potentials)} \\
				\cline{1-3}\cline{5-6}
				Spin states                                 &   $B_{\rm res}$   & $\delta$ &  &  $B_{\rm res}$ \qquad & $\Delta$ \\
				&    (G)    &   (G)    &  &   (G)  \qquad &   (G)    \\ \hline
				$\ket{\mathrm{K:a}}$+ $\ket{\mathrm{Cs:a}}$ & 361.1(1)  &  3.2(4)  &  & 341.89 &   4.7    \\
				& 442.59(1) & 0.28(3)  &  & 421.37 &   0.38   \\
				&           &          &  &        &          \\
				$\ket{\mathrm{K:b}}$+ $\ket{\mathrm{Cs:a}}$ & 419.3(1)  &  3.0(5)  &  & 399.93 &   4.3    \\
				& 513.12(1)  & 0.16(6)  &  & 491.39 &   0.55   \\
				&           &          &  &        &          \\
				$\ket{\mathrm{K:c}}$+ $\ket{\mathrm{Cs:a}}$ & 491.5(1)  &  2.1(4)  &  & 471.97 &   3.8    \\
				& 599.32(3) &  0.5(1)  &  & 575.67 &   0.44
			\end{tabular}
		\end{center}
	\end{ruledtabular}
\end{table}
%%%%%%%%%%%%%%%%%%%%%%%%%%%%%%%%%

First, we perform Feshbach spectroscopy on a
$\ket{\mathrm{K:c}}$-$\ket{\mathrm{Cs:a}}$ mixture. For this, we linearly ramp
the magnetic offset field $B$ within 10~ms to any desired value in the range
from $0$ to $650$~G and hold it there for 900 to 1300~ms. During this hold
time we exponentially decrease the power of the transformable beam to 520~mW 
to enhance the loss of K atoms from the crossed trap. In the vicinity of
an interspecies Feshbach resonance, the K sample undergoes enhanced trap loss
through three-body recombination and heating from the interaction with the
hotter Cs sample. To detect the remaining fraction of K atoms we ramp $B$
within 10~ms to 0.1~G before applying standard absorption imaging. For this
particular spin mixture we scan $B$ from 0 to 650 G in steps of 1 G and observe
two loss features, one broad and one narrow, located around 491.5 and
599.3~G, respectively. We scan the loss features with finer resolution in $B$.
The loss occurs over a range of 0.1 to several G, depending on the resonance
and the specific experimental conditions. The results around 491.5~G are shown
in Fig.~\ref{FIG2}(a). The K atom number shows a clear loss maximum. The loss
minimum that appears around 495.5~G may be the result of the zero-crossing of
the scattering length on the high-field side of the resonance. We fit
Lorentzian functions to the loss features to obtain the positions of maximum
loss $B_{\rm res}$ and the full widths at half maximum (FWHM) $\delta$.

\begin{figure*}[t]
	\begin{center}{
			\includegraphics[width=2\columnwidth]{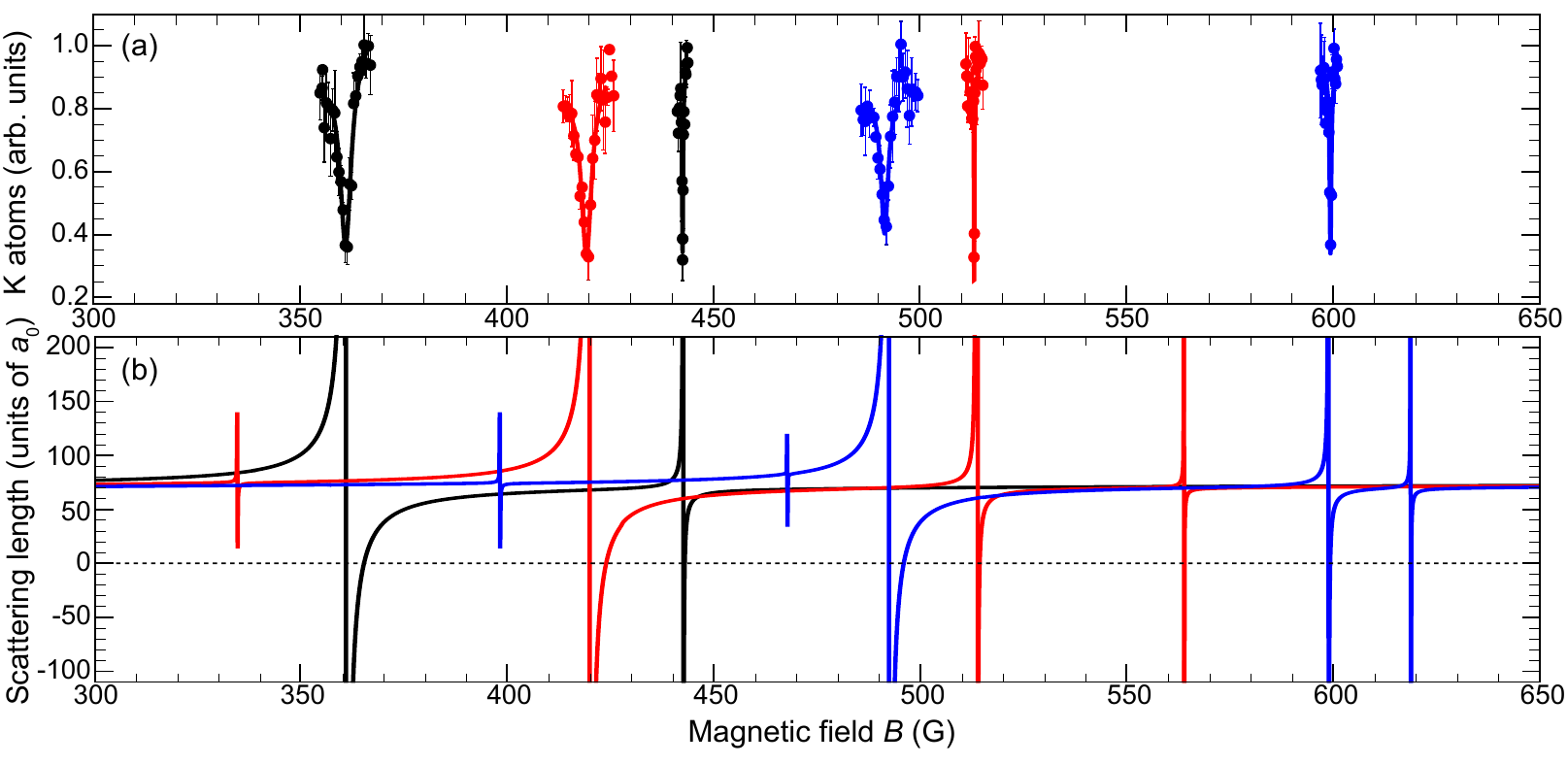}}
		\caption{\label{FIG3}Overview of $^{39}$K-$^{133}$Cs interspecies Feshbach
			resonances. (a) Loss features for $\ket{\mathrm{Cs:a}}$ with
			$\ket{\mathrm{K:a}}$ (black), $\ket{\mathrm{K:b}}$ [red (light gray)], and
			$\ket{\mathrm{K:c}}$ [blue (dark gray)]. The data are normalized to the atom number away from
			resonance. (b) Calculated interspecies scattering length $a$ for the three
			hyperfine state combinations as a function of magnetic field $B$, using the
			G2017 interaction potentials fitted in Sec.~\ref{sec:fitting}.}
	\end{center}
\end{figure*}

We also carry out time-of-flight measurements to determine the
temperature of the K sample. The results are shown in Fig.~\ref{FIG2}(b); we
observe an increase in temperature from 3.0 to 4.5~$\mu$K at the same
location as the loss is maximal. This temperature increase vanishes when the Cs
sample is absent. We attribute the increase in temperature to partial
thermalization with the hotter Cs sample. Higher temperatures are probably
counteracted by evaporation due to the finite trap depth. As will
be seen in Sec.~\ref{sec:predict} below, the background scattering length
for $^{39}$K-Cs is around 70\ $a_0$, and this relatively small value explains
the absence of observed thermalization away from resonance.

For Feshbach spectroscopy with K atoms in $\ket{\mathrm{K:b}}$ and
$\ket{\mathrm{K:a}}$ we transfer the K atoms by radio-frequency adiabatic
passage from $\ket{\mathrm{K:c}}$ to $\ket{\mathrm{K:b}}$ and, subsequently, to
$\ket{\mathrm{K:a}}$. Each step takes place at $B=35.5$~G within 25~ms with an
efficiency close to unity. For each spin mixture we again detect a pair of
resonances, one broader and the other narrower. All measured Feshbach
resonances are shown in Fig.\ \ref{FIG3}(a) and their parameters are summarized
in Table~\ref{TABLE1}. We obtain the magnetic field $B$ by measuring Cs
microwave frequencies at the fields where interspecies loss features are
observed. The experimental errors given in Table~\ref{TABLE1} are statistical
errors from the Lorentzian fits. A drift of the magnetization of the stainless
steel chamber, which depends on the offset field strength, gives rise to a
systematic error on the order of $\pm$0.3~G.

\section{Theory}

\subsection{Computational methods for bound states and scattering}

For the scattering and near-threshold bound states, we solve the Schr\"odinger
equation by coupled-channel methods, using a basis set for the electron and
nuclear spins in a fully decoupled representation,
\begin{equation}
|s_{\rm K} m_{s,{\rm K}}\rangle|i_{\rm K} m_{i,{\rm K}}\rangle
|s_{\rm Cs} m_{s,{\rm Cs}}\rangle |i_{\rm Cs} m_{i,{\rm Cs}}\rangle |L
M_L \rangle. \label{eqbasdecoup}
\end{equation}
The matrix elements of the different terms in the Hamiltonian in this basis set
are given in the Appendix of Ref.~\cite{Hutson:Cs2-note:2008}. The calculations
in this paper used basis sets with all possible values of $m_s$ and $m_i$ for
both atoms, subject to conservation of $M_{\rm tot}=m_{s,{\rm K}}+m_{i,{\rm
K}}+m_{s,{\rm Cs}}+m_{i,{\rm Cs}}+M_L$ and parity $(-1)^L$. For s-wave scattering at a particular
threshold, $M_{\rm tot}$ is set by the states of the incoming atoms, $M_{\rm
tot}=m_{f,{\rm K}}+m_{f,{\rm Cs}}$, and only channels with even $L$ contribute.

Scattering calculations are carried out using the MOLSCAT package
\cite{molscat:v14}, as modified to handle collisions in magnetic fields
\cite{Gonzalez-Martinez:2007}. At each magnetic field $B$, the wave-function
log-derivative matrix at collision energy $E$ is propagated from $R=5.6$~$a_0$
to 15~$a_0$ using the propagator of Manolopoulos \cite{Manolopoulos:1986} with
a fixed step size of 0.001~$a_0$, and from 15 to 3,000~$a_0$ using the Airy
propagator \cite{Alexander:1987} with a variable step size controlled by the
parameter $\mathrm{TOLHI} = 10^{-5}$ \cite{Alexander:1984}. Scattering boundary
conditions \cite{Johnson:1973} are applied at $R=3,000$~$a_0$ to obtain the
scattering S-matrix. The energy-dependent s-wave scattering length $a(k)$ is
then obtained from the diagonal S-matrix element in the incoming $L=0$ channel
using the identity \cite{Hutson:res:2007}
\begin{equation}
\label{scat-length}
a(k) = \frac{1}{ik} \left(\frac{1-S_{00}}{1+S_{00}}\right),
\end{equation}
where $k^2=2\mu E/\hbar^2$ and $\mu$ is the reduced mass. This
reduces to the standard zero-energy scattering length in the low-energy limit.

Weakly bound levels for Feshbach molecules are obtained using the propagation
method described in Refs.~\cite{Hutson:CPC:1994,Hutson:Cs2-note:2008}, using the same step size as for MOLSCAT with a
reduced propagation range of $R=5.6$~$a_0$ to $1,000$~$a_0$. Levels are located
either as bound-state energies at a fixed value of the magnetic field $B$ using
the BOUND package \cite{Hutson:bound:1993} or as bound-state fields at a fixed
value of the binding energy using the FIELD package \cite{Hutson:field:2011}.
BOUND and FIELD converge to values of the energy (or field) where the
log-derivative matching matrix \cite{Hutson:CPC:1994} has a zero eigenvalue.
Both programs implement a node-count algorithm \cite{Hutson:CPC:1994} which
makes it straightforward to ensure that {\em all} bound states that exist in a
particular range of energy or field are located.

Zero-energy Feshbach resonances can be located as fields $B_{\rm res}$ at which
the scattering length $a(B)$ passes through a pole,
\begin{equation}
a(B) = a_{\rm bg} \left(1-\frac{\Delta}{B-B_{\rm res}}\right).
\end{equation}
MOLSCAT has the capability to converge on such poles to provide resonance
widths $\Delta$ and background scattering lengths $a_{\rm bg}$ as well as
resonance positions $B_{\rm res}$. However, when only resonance positions are required, the
FIELD package provides a much cleaner approach: simply running FIELD at zero
energy provides a complete list of the energies at which bound states cross
threshold, and thus a complete list of resonance positions. The widths and
background scattering lengths may then be obtained if required, using
scattering calculations with MOLSCAT around the field concerned.

In the present work, basis sets including only $L=0$ functions were used in
most cases, since they make the calculations simpler at the b + a and c + a
thresholds, where inelastic decay would otherwise exist. However, calculations
with $L_{\rm max}=2$ were used for the calculations of scattering and bound
states on the fitted potentials in Sec.~\ref{sec:predict} below. As will be
seen, the observed resonances in the a + a channel shifted by no more than 0.01~G
when $L=2$ basis functions were included, which is considerably less than the
experimental uncertainties in the resonance positions.

\subsection{Potential curves}

The KCs interaction potentials of Ferber {\em et al.}\ \cite{Ferber2013}
(F2013) were fitted to extensive Fourier transform spectra of the KCs molecule,
including vibrational levels up to $v=102$ for the $X^1\Sigma^+$ singlet ground
state and $v=32$ for the $a^3\Sigma^+$ triplet state (although there is
significant mixing of the singlet and triplet states for the highest
vibrational levels). Each potential curve is constructed in three segments; the
central segment from $R^{\rm SR}_S$ to $R^{\rm LR}_S$, with $S=0$ or 1 for the
singlet or triplet state, respectively, is represented as a power-series
expansion in the variable $\xi(R)= (R-R_{\rm m})/(R+bR_{\rm m})$, where $R_{\rm
m}$ is chosen to be near the equilibrium distance. At long range ($R
> R^{\rm LR}_S$), the potentials are
\begin{equation}
\begin{split}
V_S^{\rm LR}(R) = -C_6/R^6 - C_8/R^8 - C_{10}/R^{10}\\
-(-1)^S V_{\rm exch}(R),
\end{split}
\end{equation}
where the dispersion coefficients $C_n$ \cite{Derevianko:2001, Porsev:2003,
Ferber2013} are common to both potentials. The exchange contribution is
\cite{Smirnov:1965}
\begin{equation}
V_{\rm exch}(R) = A_{\rm ex} R^\gamma \exp(-\beta R),
\end{equation}
and makes an attractive contribution for the singlet and a repulsive
contribution for the triplet. The central segment is constrained to match the
long-range potential at $R_S^{\rm LR}$. The potentials are extended to short
range ($R < R_S^{\rm SR}$) with simple repulsive terms,
\begin{equation}
V_S^{\rm SR}(R) = A^{\rm SR}_S + B^{\rm SR}_S / R^{N^{\rm SR}_S}.
\end{equation}
The parameters $A^{\rm SR}_S$ and $B^{\rm SR}_S$ are chosen to match the values
and derivatives of the mid-range potentials at $R_S^{\rm SR}$. The potential
matching points for KCs are $R_0^{\rm SR}=3.22$~\AA\ and $R_0^{\rm
LR}=12.00$~\AA\ for the singlet state and $R_1^{\rm SR}=5.23$~\AA\ and
$R_1^{\rm LR}=12.01$~\AA\ for the triplet state \cite{Ferber2013}.

\begin{table}[t]
	\caption{Quality of fit to the observed resonance positions, together with the
		properties of additional resonances due to s-wave bound states predicted by the G2017 potentials. The
		uncertainties given here include systematic errors and are those that define
		the weights used in the least-squares fit. \label{quality-of-fit}}
	\begin{tabular}{clcllcrcr}
		\hline\hline
		\omit\hfill\vrule height 2ex depth 1ex width 0pt Threshold & $B_{\rm obs}$\hfill&\qquad& $B_{\rm calc}$& $\Delta_{\rm calc}$\hfill&\multispan3\hfill$B_{\rm obs}-B_{\rm calc}$\hfill & Unc. \\
		\hline
		a + a & 361.1   && 360.74\qquad\qquad & 4.4   &&    0.36 && 0.4 \\
		a + a & 442.59  && 442.43 & 0.37  &&    0.16 && 0.3 \\
		\\
		b + a & 419.3   && 419.73 & 4.0   && $-$0.43 && 0.4 \\
		b + a & 513.12  && 513.73 & 0.52  && $-$0.61 && 0.3 \\
		\\
		c + a & 491.5   && 492.24 & 3.6   && $-$0.74 && 0.4 \\
		c + a & 599.32  && 598.76 & 0.39  &&    0.58 && 0.3 \\
		\\
		b + a &         && 334.45 & 0.025 \\
		b + a &         && 563.81 & 0.074 \\
		\\
		b + b &         && 319.35 & 0.046 \\
		c + a &         && 398.09 & 0.023 \\
		c + a &         && 467.70 & 0.006 \\
		c + a &         && 618.71 & 0.094 \\
		
		\hline\hline
	\end{tabular}
\end{table}

For coupled-channel calculations of the near-threshold bound states and
scattering properties, these potentials are supplemented by a coupling $\hat
V^{\rm d}(R)$, which at long range has a simple magnetic dipole-dipole form
that varies as $1/R^3$~\cite{Stoof:1988, Moerdijk:1995}. However, for heavy
atoms, second-order spin-orbit coupling provides an additional contribution
that has the same tensor form as the dipole-dipole term. $\hat V^{\rm d}(R)$ is
represented as
\begin{equation}
\label{eq:Vd} \hat V^{\rm d}(R) = \lambda(R) \left [ \hat s_1\cdot
\hat s_2 -3 (\hat s_1 \cdot \vec e_R)(\hat s_2 \cdot \vec e_R)
\right ] \,,
\end{equation}
where $\vec e_R$ is a unit vector along the internuclear axis and $\lambda$ is
an $R$-dependent coupling constant,
\begin{eqnarray}
\label{eq:lambda}
\lambda(R) &=& E_{\rm h} \alpha^2 \bigg[
A_{\rm 2SO}^{\rm short} \exp\left(-\beta_{\rm 2SO}^{\rm short}(R/a_0)\right)
\nonumber\\
&+& A_{\rm 2SO}^{\rm long}
\exp\left(-\beta_{\rm 2SO}^{\rm long}(R/a_0)\right)
+  \frac{g_S^2}{4(R/a_0)^3}\bigg],
\end{eqnarray}
where $\alpha\approx 1/137$ is the atomic fine-structure constant, $E_{\rm h}$
is the Hartree energy and $g_S\approx2.0023$ is the electron $g$-factor. The
second-order spin-orbit coupling has not been obtained from electronic
structure calculations for KCs, so in the present work we retained the estimate
used in Ref.\ \cite{Patel2014}, obtained by shifting the RbCs function
\cite{Takekoshi:RbCs:2012} inwards by 0.125~$a_0$, to give the same value at
the inner turning point for KCs as for RbCs. This gives $\beta_{\rm 2SO}^{\rm
short} = 0.80$ and $\beta_{\rm 2SO}^{\rm long} = 0.28$ as for RbCs, with
$A_{\rm 2SO}^{\rm short} = -45.5$ and $A_{\rm 2SO}^{\rm long} = -0.032$.

\begin{figure}[t]
	\begin{center}{
			\includegraphics[width=\columnwidth,clip=true,trim=0.5cm 1.9cm 1.9cm 2.8cm]{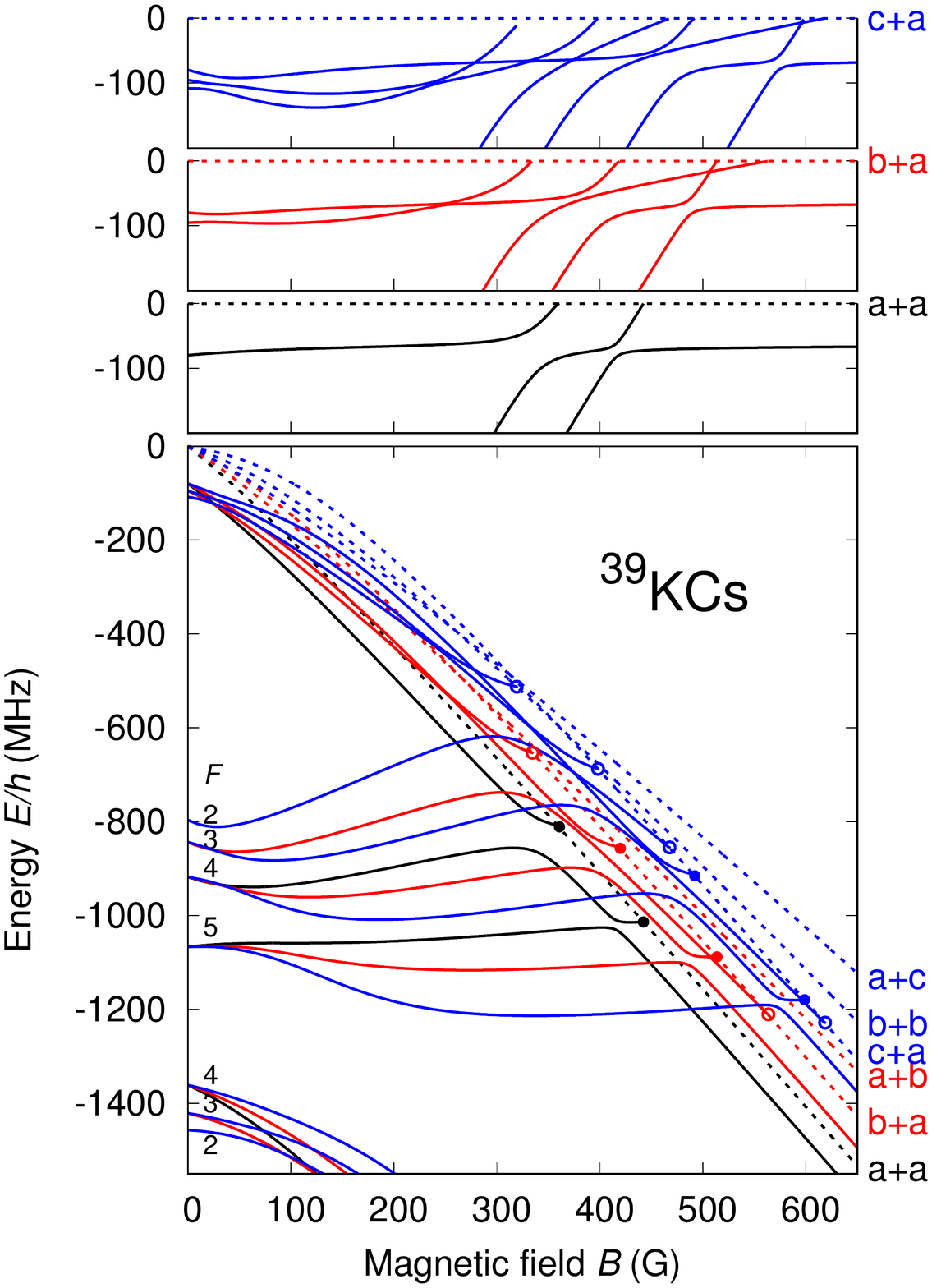}}
\caption{Bottom panel: Bound states of $^{39}$KCs (solid lines) with $M_F=2$
[blue (dark gray)], 3 [red (light gray)] and 4 (black), together with the corresponding thresholds
(dashed lines), calculated using the G2017 fitted interaction potentials of
Sec.~\ref{sec:fitting} with $L=0$ functions only and shown with respect to
the lowest zero-field threshold. The threshold crossings that produce observed
resonances are shown with filled circles and those so far unobserved with open
circles. Upper panels: Bound states for $M_F=2$, 3 and 4, shown relative to the field-dependent
c + a, b + a and a + a thresholds. \label{bound-allM}}
	\end{center}
\end{figure}

\subsection{Least-squares fitting}
\label{sec:fitting}

The resonances observed in the present work are due to s-wave bound states that
cross the threshold as a function of magnetic field. These are substantially
broader than resonances due to d-wave and higher states \cite{Patel2014}, 
which appear only because of the weak anisotropic term $\hat V^{\rm d}(R)$ in the Hamiltonian.
Figure \ref{bound-allM} shows the atomic thresholds for $M_F=m_{f,{\rm
K}}+m_{f,{\rm Cs}}=2$, 3 and 4, together with the s-wave bound states
responsible for the resonances observed here, calculated on the G2017 fitted
potentials described below. There is a state roughly parallel to each of the
a + a, b + a and c + a thresholds, bound by 65 to 70~MHz and with the same ($f_{\rm
K},m_{f,{\rm K}},f_{\rm Cs},m_{f,{\rm Cs}}$) character as the threshold.
Crossing these near-threshold states are a set of deeper states, bound by 800
to 1100~MHz at zero field, that are closer to horizontal in the bottom panel of
Fig.~\ref{bound-allM}. This second set of states correlates at zero field with
atoms with $f_{\rm K}=2$ and $f_{\rm Cs}=3$, with resultant $F=1$, 2, 3, 4, 5
(though $F=1$ does not appear in Fig.~\ref{bound-allM} because states with
$M_F<2$ are not shown). The observed resonances occur when the near-threshold
states are pushed across threshold by mixing with the near-horizontal states at
broad avoided crossings, which are complete to varying extents at threshold.
The positions of the resonances are thus principally determined by the binding
energies of the near-horizontal states, although the states actually crossing
threshold have mixed character. An analogous figure showing the
bound states on the F2013 potentials \cite{Ferber2013} is included in the
Supplemental Material \cite{suppl}; the pattern of states is visually very similar, despite
that fact that the sign of $a_{\rm s}-a_{\rm t}$ is reversed on the F2013 potentials.

The right-hand columns of Table \ref{TABLE1} give the calculated positions and
widths of the observed resonances, obtained using the F2013 potentials. It may
be seen that the calculated resonance positions are all about 20~G lower than the
experiment. It is therefore desirable to adjust the interaction potentials to
reproduce the resonance positions. In doing this, we wish to retain as much as
possible of the spectroscopically determined potentials of Ref.\
\cite{Ferber2013}, so that the fit to the Fourier transform spectra is affected
as little as possible. We found in the initial fitting that it is possible to
reproduce the scattering properties by retaining the central and long-range
parts of the spectroscopic potential curves and adjusting only the short-range
parts for $R < R_S^{\rm SR}$. Small changes to the potential curves in this
region have relatively little effect on levels with inner turning points below
$V_0(R_0^{\rm SR})/hc=-316.6$~cm$^{-1}$ for the singlet state and $V_1(R_1^{\rm
SR})/hc=-116.1$~cm$^{-1}$ for the triplet state. We explored modifications to
the values of $N_S^{\rm SR}$, with corresponding changes in $A_S^{\rm SR}$ and
$B_S^{\rm SR}$ to match the values and derivatives of the power-series
expansions at $R = R_S^{\rm SR}$,
\begin{eqnarray}
B_S^{\rm SR} &=& -\left(\frac{(R_S^{\rm SR})^{N_S^{\rm SR}+1}}{N_S^{\rm SR}}\right)
\left(\frac{dV_S}{dR}\right)_{R=R_S^{\rm SR}};\nonumber\\
A_S^{\rm SR} &=& V_S(R_S^{\rm SR}) - B_S^{\rm SR}/(R_S^{\rm SR})^{N_S^{\rm SR}}.\label{eq:absr}
\end{eqnarray}

\begin{table}[tbp]
	\caption{Calculated singlet and triplet scattering lengths for isotopologs of
		KCs, with 1-$\sigma$ statistical uncertainties for $a_{\rm t}$.}
	\label{tab:scat-len}
	\begin{tabular}{lrrr}
		\hline\hline
		&$a_{\rm s}$ ($a_0$)&$a_{\rm t}$ ($a_0$) & $a_{\rm t}$ ($a_0$)\\
		&       &G2017 potentials &F2013 potentials\\
		\hline
		$^{39}$KCs & $-$18.37 &    74.88(9)\ &    82.24 \\
		$^{40}$KCs & $-$51.44 & $-$71.67(45) & $-$41.28 \\
		$^{41}$KCs & $-$72.79 &   179.06(28) &   205.25 \\
		\hline\hline
	\end{tabular}
\end{table}

\begin{figure*}[t]
	\includegraphics[width=2.05\columnwidth]{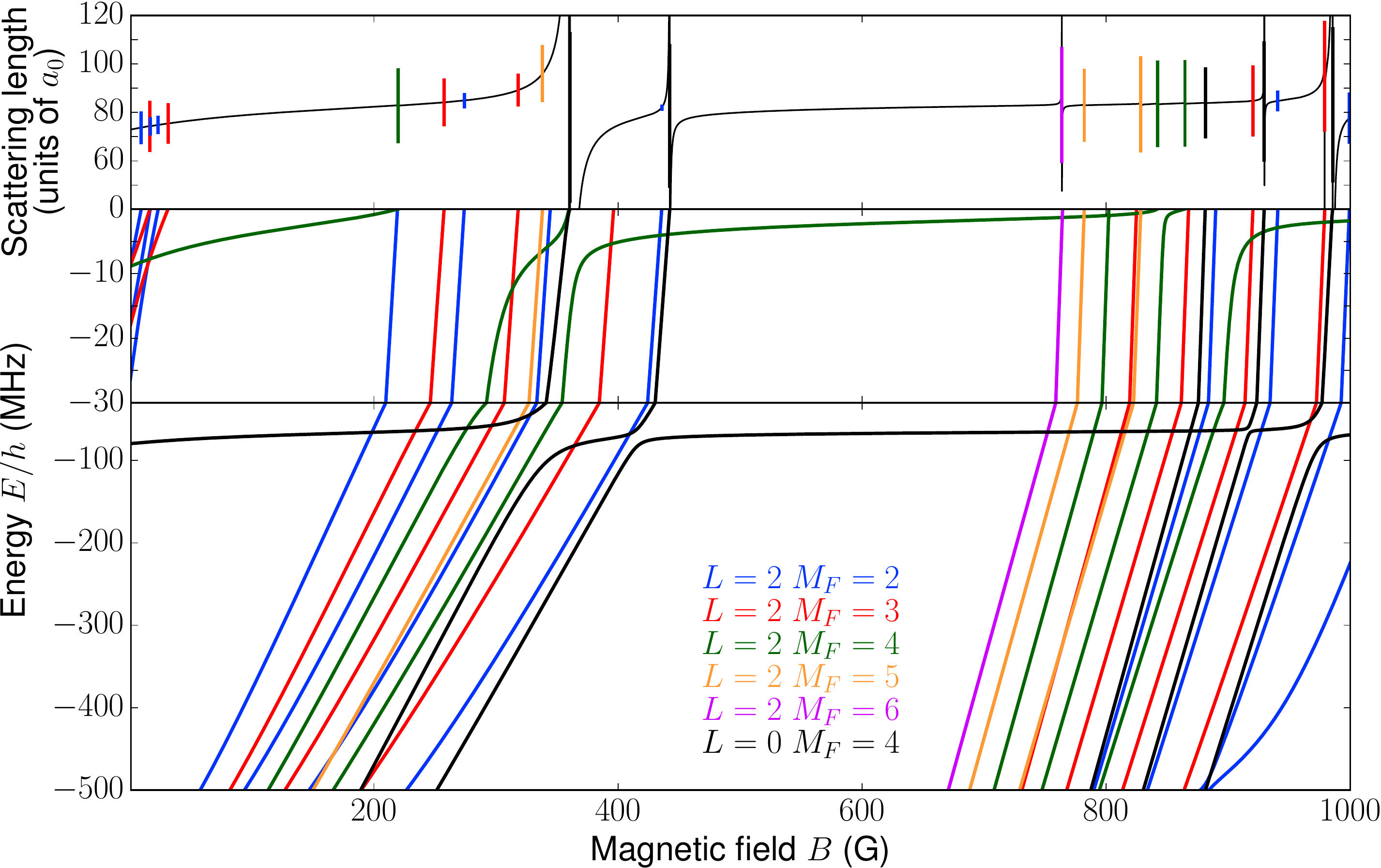}
	\caption{Scattering length in the a + a channel and energies of near-threshold
		bound states for $^{39}$KCs. Note that the top 30 MHz is shown on an expanded
		scale. Resonance widths greater than 1~$\mu$G are shown as vertical bars with
		lengths proportional to $\log (\Delta/\mu$G). \label{fig:predict39}}
\end{figure*}

We carried out least-squares fits of potential parameters to the observed
resonance positions using the Interactive Non-Linear Least-Squares (I-NoLLS) package \cite{I-NoLLS}, which gives the user interactive control over step
lengths and assignments as the fit proceeds. The quantity optimized in the
least-squares fits was the sum of squares of residuals [(observed $-$
calculated)/uncertainty], with the uncertainties listed in Table~
\ref{quality-of-fit}. The resonance positions at the a + a, b + a and c + a
thresholds are principally sensitive to the triplet potential, so our initial
fits varied $N_1^{\rm SR}$ only, leaving the singlet potential unchanged. The
optimum fit was obtained with $N_1^{\rm SR}=6.9(1)$ and gives the fit to the
resonance positions shown in Table~\ref{quality-of-fit}. The short-range
coefficients for the best-fit triplet potential are obtained from Eq.\
(\ref{eq:absr}); their approximate values are $A_1^{\rm SR}/hc\approx -471.5728$
cm$^{-1}$ and $B_1^{\rm SR}/hc\approx 3.224735 \times 10^7$~cm$^{-1}$~\AA$^{N_1^{\rm SR}}$.

We subsequently explored two-parameter fits, varying both $N_0^{\rm SR}$ and
$N_1^{\rm SR}$. However, these produced no significant improvement in the
quality of fit. A single-parameter fit varying only $N_0^{\rm SR}$ was
incapable of reproducing the observed resonance positions. We therefore decided
to proceed with the single-parameter fit obtained by varying only the triplet
potential, which we designate the G2017 potentials; a more
extensive refinement will require additional experimental results.

It may be noted that the present modification shifts the vibrational levels of
the $a^3\Sigma^+$ state by a maximum of about 0.1~cm$^{-1}$; this shift is less
than 0.01~cm$^{-1}$ below $v=11$ and peaks around $v=22$. These shifts are
smaller than the Cs hyperfine splitting, but larger than the typical
experimental uncertainties of 0.01~ cm$^{-1}$ in Ref.\ \cite{Ferber2013}. A
more complete treatment would require refitting the entire potentials, but is
not justified at this stage.

\begin{figure*}[tbp]
\includegraphics[width=2.05\columnwidth]{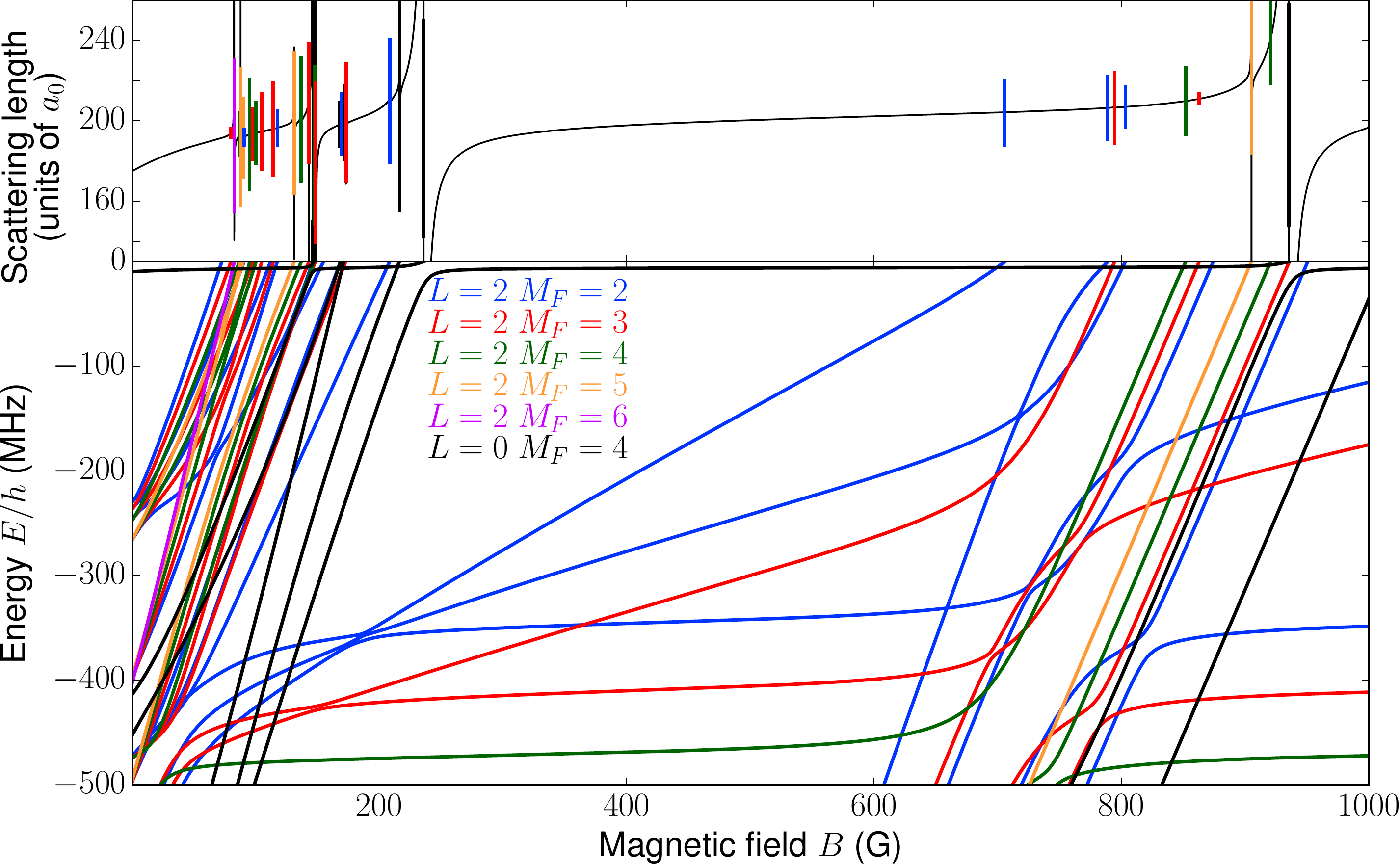}
\caption{Scattering length in the a + a channel and energies of near-threshold
bound states for $^{41}$KCs. Resonance widths greater than 1~$\mu$G are shown
as vertical bars with lengths proportional to $\log (\Delta/\mu$G).
\label{fig:predict41}}
\end{figure*}

\begin{figure*}[tbp]
\includegraphics[width=2.05\columnwidth]{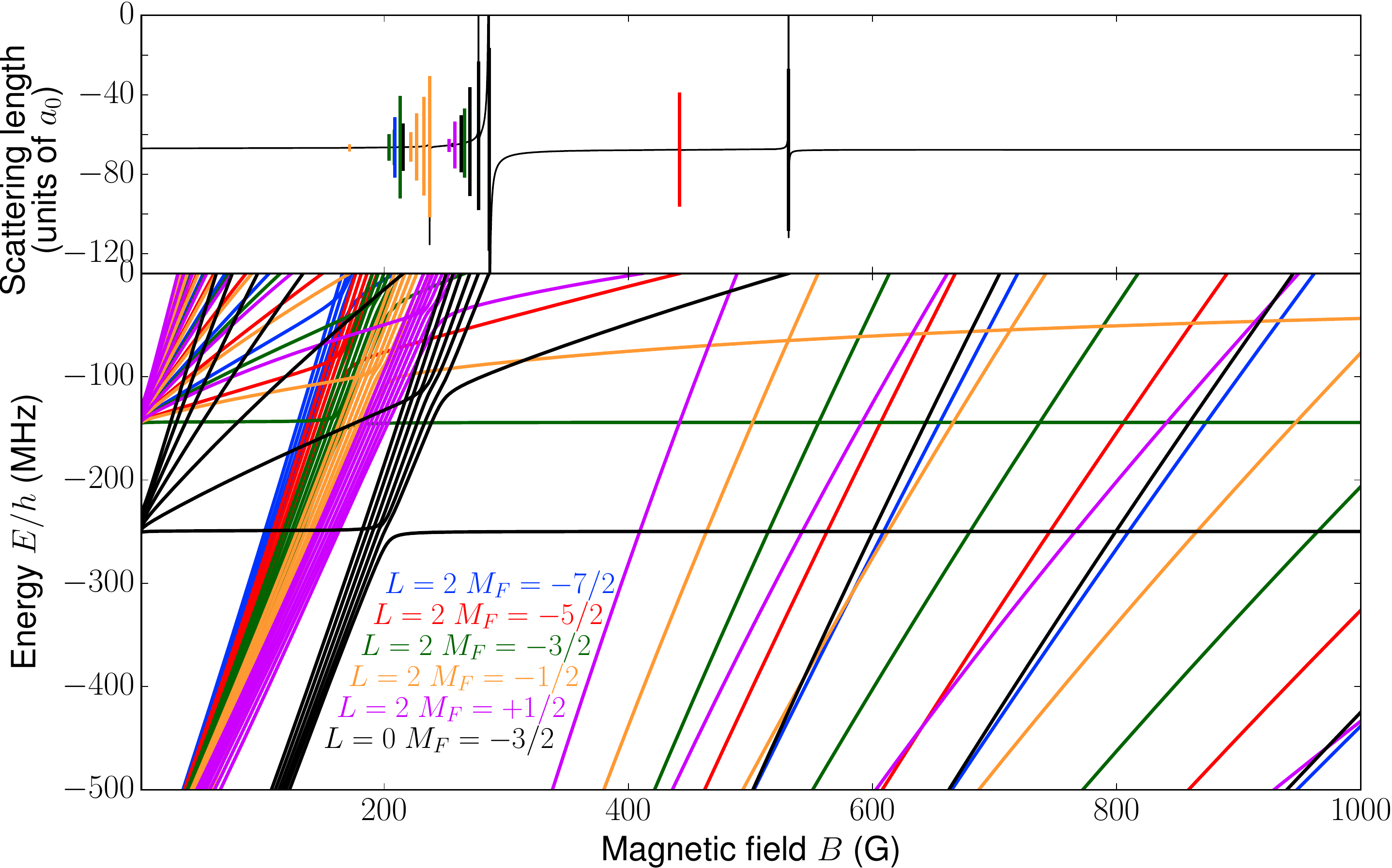}
\caption{Scattering length in the a + a channel and energies of near-threshold
bound states for $^{40}$KCs. Resonance widths greater than 1~$\mu$G are shown
as vertical bars with lengths proportional to $\log (\Delta/\mu$G).
\label{fig:predict40}}
\end{figure*}

\subsection{Calculations on optimized potentials}\label{sec:predict}

Figures \ref{fig:predict39}, \ref{fig:predict41}, and \ref{fig:predict40} show the scattering lengths
and resonance positions, calculated at the lowest (a + a) threshold for all three
isotopic combinations of KCs using the optimized G2017 potentials, together
with the near-threshold bound states that produce the resonances.
Table~\ref{tab:scat-len} gives the predicted singlet and triplet scattering
lengths $a_{\rm s}$ and $a_{\rm t}$, together with the statistical uncertainty
of $a_{\rm t}$ in the one-parameter space. However, $a_{\rm t}$ may change
outside these limits when additional parameters are fitted. Complete lists of
the resonance parameters (positions, widths and background scattering lengths)
are given in the Supplemental Material \cite{suppl}.

The G2017 potentials obtained here predict scattering and bound-state
properties that differ in some important ways from those of Ref.\
\cite{Patel2014}. In particular, the triplet scattering lengths all shift to
smaller (or more negative) values, and the corresponding near-threshold levels
are more deeply bound.

Figure \ref{bound-allM} shows that, in addition to the two
resonances currently observed in each of the a + a, b + a and c + a channels, there
are additional $^{39}$KCs resonances due to s-wave states in the b + a, c + a and
b + b channels (where b + b is the lowest $M_F=2$ threshold at fields between 261~G
and 358~G). The positions and widths of these are included in Table
\ref{quality-of-fit}. They have predicted widths between 6~mG and 0.1~G, and
may be useful for molecule formation.

For $^{39}$KCs, Patel {\em et al.}\ \cite{Patel2014} predicted a group of low-field
resonances at fields below 50~G, due to a group of d-wave states bound by less
than 20~MHz at zero field. The widest of these was a resonance predicted near
50~G with a width of 1~mG, due to a state that was bound by less
than 3~MHz at zero field but was nearly parallel to threshold as a function of
field. On the G2017 potentials the corresponding states are significantly
deeper; the state responsible for the widest resonance is now bound by about
9~MHz at zero field, and the resulting resonance is shifted to 219~G, now with
a width of 0.7~mG. Similarly, for $^{41}$KCs, Patel {\em et al.}\ \cite{Patel2014} predicted
a group of low-field resonances at fields of 20 to 30~G, again due to a group
of very weakly bound d-wave states. These looked promising for molecule
formation because of their proximity to the region around 21~G where Cs can be
cooled to degeneracy. On the G2017 potentials, however, these states are again
significantly deeper, and the resonances are shifted to fields above 70~G.

For fermionic $^{40}$KCs, by contrast, the G2017 potentials appear to offer
improved prospects for molecule formation. The older F2013 potentials
\cite{Ferber2013} give scattering lengths $a_{\rm s}=-51.44\ a_0$ and $a_{\rm
t}=-41.28\ a_0$. Because these are so similar, even resonances due to s-wave
states were predicted in Ref.\ \cite{Patel2014} to be very narrow and those due
to d-wave states even narrower. The G2017 potential, however, has $a_{\rm
t}=-71.67\ a_0$. Because of this, the resonances are shifted to rather higher
fields, but they are also considerably broader. For example, the resonance
predicted in Ref.\ \cite{Patel2014} at 264.3~G with a width of $-0.1$~G occurs
at 286.0~G on the G2017 potentials with a width of $-0.86$~G. Similarly, the
resonance previously predicted at 470~G with a width of $-10$~mG is now at
531~G, with a width of $-54$~mG. The latter is particularly promising for
molecule formation, because it is reasonably close to the region around 556~G
where Cs has a moderate scattering length \cite{Berninger2013} and can be
cooled efficiently.

\section{Conclusion}
We observed Feshbach resonances due to s-wave bound states in ultracold
collisions of $^{39}$K and Cs. The resonances occur at magnetic fields about
20~G higher than those predicted in Refs.\ \cite{Ferber2013,Patel2014} using interaction potentials fitted to high-resolution Fourier
transform spectra. Reproducing the experimental resonance positions requires a
significant change to the triplet interaction potential found in Ref.\
\cite{Ferber2013}. We carried out least-squares fits to determine a
triplet potential with a modified repulsive wall, which reproduces the Feshbach
resonance positions while making only small changes to the deeper vibrational
levels.

We used the modified interaction potentials, which we
designate G2017, to carry out coupled-channels calculations and make improved
predictions of the near-threshold bound states and ultracold scattering
properties for all three isotopes of K interacting with Cs. For the case of
$^{40}$KCs, the scattering properties are more favorable using the G2017
potentials than was found in Refs.\ \cite{Ferber2013,Patel2014}. In
particular, the G2017 potentials predict a Feshbach resonance that is broad
enough to allow tuning of the interactions in a K-Cs Fermi-Bose mixture. The
results open up various interesting avenues in cold atom and cold molecule
research. These include studies of the dynamics and transport properties of
bosonic impurities in low-dimensional Fermi gases, similar to recent
experiments where Bloch-type oscillations have been observed for impurity
motion in a fermionized one-dimensional Bose gas \cite{Meinert2016boi}. It may also be
possible to form fermionic KCs molecules and transfer them to the rovibrational
ground state to generate dipolar quantum gases, employing techniques such as
those recently demonstrated for fermionic KRb \cite{Moses2015} and for bosonic
RbCs \cite{Reichsollner2016}.
\\
\\
\section{Acknowledgments}

We are indebted to R. Grimm for generous support. We thank G. Anich, K.
Jag-Lauber, F. Meinert, and G. Unnikrishnan for fruitful discussions. 
A subroutine to evaluate the potential curves developed in this work, 
together with sample input and output files for the MOLSCAT, BOUND and 
FIELD programs and the associated experimental data, has been deposited at doi:10.15128/r1cf95jb44k.
We gratefully acknowledge funding by the European Research Council (ERC) under
Project No.\ 278417, by the Austrian Science Foundation (FWF) under Project
No.\ I1789-N20 (joint Austrian-French FWF-ANR project) and under Project No.\
P29602-N36 and by the UK Engineering and Physical Sciences Research Council
under Grant No.\ EP/N007085/1.

\bibliographystyle{apsrev}
\bibliography{KCsFeshbach,ultracold,ultracold_molecules_HCN,all}

\newpage
\clearpage

\onecolumngrid
\section{Supplemental Material: Observation of interspecies Feshbach resonances in an ultracold $^{39}$K-$^{133}$Cs mixture and refinement of interaction potentials}
\twocolumngrid

\begin{table*}[tbp] 
	\caption{Resonance properties predicted from the
		optimized G2017 potentials for the a + a channel of $^{39}$KCs, with the
		positions from Ref.\ \cite{Patel2014} for comparison. Resonances were located
		using MOLSCAT, with $R_{\rm max}$ set to 1,500\ $a_0$.} \label{tab:predict39}
	
	\begin{tabular}{rcrccrrcrr}
		\hline\hline
		$B_{\rm res}$ (G) &\qquad& $B_{\rm res}$ (G)&\qquad& $\Delta$ (G) & \multispan3\hfill $a_{\rm bg}$ ($a_0$) \hfill& $L$ & $M_F$\\
		G2017&& F2013 && G2017 && G2017 \hfill &&\qquad\\
		potentials && potentials &&potentials&&potentials&&\qquad\\
		\hline
		9.335 &&   2.995  &&  $1.5\times10^{-5}$   &&   63.7  && 2 & 2 \\
		16.450 &&   5.482  &&  $8\times10^{-5}$     &&   64.3  && 2 & 3 \\
		16.670 &&  10.784  &&  $4\times10^{-6}$     &&   64.3  && 2 & 2 \\
		23.293 &&  16.311  &&  $4\times10^{-6}$     &&   64.9  && 2 & 2 \\
		31.344 &&  18.154  &&  $3\times10^{-5}$     &&   65.5  && 2 & 3 \\
		219.346 && 202.900  &&  $6\times10^{-7}$     &&   72.9  && 2 & 2 \\
		219.972 && 49.593   &&  $7\times10^{-4}$     &&   72.8  && 2 & 4 \\
		257.539 && 240.221  &&  $6\times10^{-5}$     &&   74.2  && 2 & 3 \\
		274.019 && 255.726  &&  $3\times10^{-6}$     &&   74.9  && 2 & 2 \\
		318.249 && 298.576  &&  $1.5\times10^{-5}$   &&   79.2  && 2 & 3 \\
		338.197 && 318.143  &&  $1.4\times10^{-4}$   &&   86.0  && 2 & 5 \\
		344.508 && 323.897  &&  $6\times10^{-7}$     &&   91.5  && 2 & 2 \\
		359.899 && 326.943  &&   0.012               &&   420   && 2 & 4 \\
		360.745 && 341.895  &&   4.6                 &&   68    && 0 & 4 \\
		396.419 && 373.849  &&  $3\times10^{-7}$     &&   63.8  && 2 & 3 \\
		435.855 && 412.433  &&  $1.3\times10^{-6}$   &&   72.0  && 2 & 2 \\
		442.429 && 421.364  &&   0.37                &&   68.6  && 0 & 4 \\
		763.737 && 697.020  &&   0.031               &&   73.0  && 2 & 6 \\
		782.083 && 714.609  &&  $6\times10^{-4}$     &&   73.0  && 2 & 5 \\
		802.159 && 375.354  &&  $5\times10^{-7}$     &&   73.2  && 2 & 4 \\
		824.886 && 757.456  &&  $4\times10^{-10}$    &&   73.5  && 2 & 3 \\
		828.375 && 760.131  &&   0.005               &&   73.4  && 2 & 5 \\
		842.083 && 734.709  &&   0.002               &&   73.5  && 2 & 4 \\
		864.584 && 778.978  &&   0.002               &&   73.7  && 2 & 4 \\
		867.490 && 798.340  &&  $7\times10^{-8}$     &&   73.7  && 2 & 3 \\
		881.298 && 813.139  &&  $5\times10^{-4}$     &&   73.9  && 0 & 4 \\
		889.799 && 819.977  &&  $7\times10^{-8}$     &&   74.0  && 2 & 2 \\
		920.147 && 849.801  &&  $5\times10^{-4}$     &&   74.8  && 2 & 3 \\
		929.603 && 860.524  &&   0.045               &&   74.6  && 0 & 4 \\
		940.700 && 869.412  &&  $5\times10^{-6}$     &&   74.7  && 2 & 2 \\
		978.997 && 907.538  &&   0.019               &&   84.9  && 2 & 3 \\
		985.676 && 915.564  &&   1.1                 &&   73.1  && 0 & 4 \\
		999.386 && 926.772  &&  $8\times10^{-5}$     &&   67.7  && 2 & 2 \\
		\hline\hline
	\end{tabular}
\end{table*}

\begin{table*}[tbp]
	\caption{Resonance properties predicted from the
		optimized G2017 potentials for the a + a channel of $^{41}$KCs, with the
		positions from Ref.\ \cite{Patel2014} for comparison. Resonances were located
		using MOLSCAT, with $R_{\rm max}$ set to 1,500\ $a_0$.} \label{tab:predict41}
	
	\begin{tabular}{rcrccrrcrr}
		\hline\hline
		$B_{\rm res}$ (G) &\qquad& $B_{\rm res}$ (G)&\qquad& $\Delta$ (G) & \multispan3\hfill $a_{\rm bg}$ ($a_0$) \hfill& $L$ & $M_F$\\
		G2017&& F2013 && G2017 && G2017 \hfill &&\qquad\\
		potentials && potentials &&potentials&&potentials&&\qquad\\
		\hline
		72.950  &&  50.848  &&  $ 6\times10^{-7} $      &&   171.5  && 2 & 2 \\
		80.281  &&  47.266  &&  $ 1.8\times10^{-6} $    &&   174.0  && 2 & 3 \\
		82.902  &&  23.890  &&  $ 0.029 $               &&   172.5  && 2 & 6 \\
		87.111  &&  53.035  &&  $ 1.8\times10^{-5} $    &&   173.2  && 2 & 4 \\
		88.120  &&  25.677  &&  $ 0.010 $               &&   171.9  && 2 & 5 \\
		90.238  &&  39.244  &&  $ 2\times10^{-4} $      &&   171.7  && 2 & 5 \\
		90.866  &&  56.608  &&  $ 3\times10^{-6} $      &&   171.9  && 2 & 2 \\
		95.460  &&  42.860  &&  $ 1.7\times10^{-3} $    &&   173.1  && 2 & 4 \\
		98.090  &&  55.545  &&  $ 3\times10^{-5} $      &&   173.5  && 2 & 3 \\
		100.577  &&  28.413  &&  $ 6\times10^{-5} $      &&   173.9  && 2 & 4 \\
		105.056  &&  32.104  &&  $ 1.7\times10^{-4} $    &&   174.6  && 2 & 3 \\
		112.269  &&  68.538  &&  $ 2\times10^{-8} $      &&   175.6  && 2 & 2 \\
		114.513  &&  64.875  &&  $ 5\times10^{-4} $      &&   176.0  && 2 & 3 \\
		117.826  &&  36.635  &&  $ 1.0\times10^{-5} $    &&   176.5  && 2 & 2 \\
		131.464  &&  90.098  &&  $ 0.014 $               &&   179.1  && 2 & 5 \\
		137.038  &&  98.542  &&  $ 0.004 $               &&   180.7  && 2 & 4 \\
		143.299  && 108.839  &&  $ 0.003 $               &&   188.8  && 2 & 3 \\
		146.325  &&  98.503  &&  $ 0.051   $             &&   345    && 2 & 2 \\
		146.492  && 113.926  &&  $ 0.15    $             &&    86.2  && 0 & 4 \\
		148.031  &&  87.381  &&  $ 0.044 $               &&   167.8  && 2 & 4 \\
		149.012  &&  94.276  &&  $ 0.048 $               &&   159.2  && 2 & 3 \\
		155.171  && 126.591  &&  $ 9\times10^{-7}   $    &&   173.0  && 2 & 2 \\
		168.004  && 120.590  &&  $ 1.9\times10^{-5} $    &&   178.1  && 0 & 4 \\
		169.953  &&  91.559  &&  $ 6\times10^{-5} $      &&   178.6  && 2 & 2 \\
		171.823  && 111.038  &&  $ 1.5\times10^{-4} $    &&   179.2  && 0 & 4 \\
		173.280  &&  90.442  &&  $ 0.003 $               &&   179.1  && 2 & 3 \\
		208.658  && 109.860  &&  $ 0.004 $               &&   189.9  && 2 & 2 \\
		216.655  && 171.198  &&  $ 0013 $                &&   195.8  && 0 & 4 \\
		236.047  && 168.192  &&  $ 2.2    $              &&   176    && 0 & 4 \\
		705.820  && 629.690  &&  $ 8\times10^{-5} $      &&   184.2  && 2 & 2 \\
		789.035  && 737.859  &&  $ 7\times10^{-5} $      &&   186.4  && 2 & 2 \\
		794.615  && 746.956  &&  $ 1.2\times10^{-4} $    &&   186.6  && 2 & 3 \\
		803.232  && 755.110  &&  $ 1.5\times10^{-5} $    &&   186.9  && 2 & 2 \\
		852.093  && 806.405  &&  $ 9\times10^{-5} $      &&   189.8  && 2 & 4 \\
		862.873  && 818.335  &&  $ 2\times10^{-6} $      &&   190.9  && 2 & 3 \\
		874.210  && 830.925  &&  $ 4.4\times10^{-9} $    &&   192.4  && 2 & 2 \\
		905.078  && 861.029  &&  $ 2.8\times10^{-2} $    &&   201.7  && 2 & 5 \\
		920.652  && 877.705  &&  $ 4.3\times10^{-4} $    &&   220.8  && 2 & 4 \\
		935.245  && 884.925  &&  $ 3.0    $              &&   183    && 0 & 4 \\
		935.745  && 894.133  &&  $-1.8\times10^{-5} $    &&  -909    && 2 & 3 \\
		950.626  && 910.602  &&  $ 1.6\times10^{-10}$    &&   148.3  && 2 & 2 \\
		\hline\hline
	\end{tabular}
\end{table*}

\begin{figure*}
	\begin{center}{
			\includegraphics[width=8.9cm,clip=true,trim=0.5cm 1.9cm 1.9cm 2.8cm]{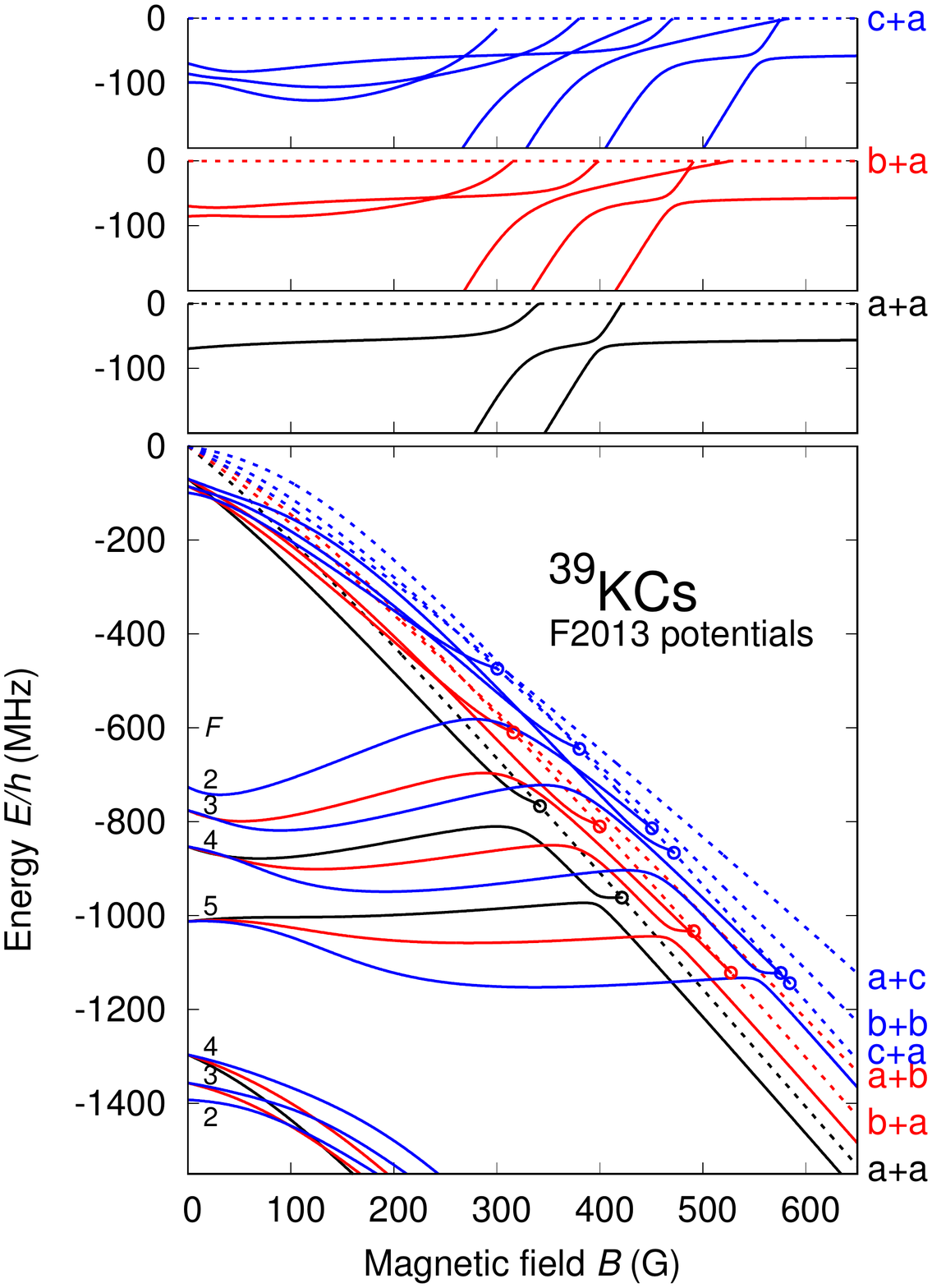}}
		\caption{Bottom panel: Bound states
			of $^{39}$KCs (solid lines) with $M_F=2$ [blue (dark gray)], 3 [red (light gray)] and 4 (black),
			together with the corresponding thresholds (dashed lines), calculated using the
			interaction potentials of Ref.\ \cite{Ferber2013} with $L=0$ functions only and
			shown with respect to the lowest zero-field threshold. Upper panels: Bound states for $M_F=2$, 3 and 4, shown relative to the field-dependent c + a, b + a
			and a + a thresholds.}
	\end{center}
\end{figure*}

\begin{table*}[tbp]
	\caption{Resonance properties predicted from the
		optimized potentials for the a + a channel of $^{40}$KCs, with the positions from
		Ref.\ \cite{Patel2014} for comparison. Resonances with widths tabulated were
		located using MOLSCAT, with $R_{\rm max}$ set to 1,500\ $a_0$; those without
		tabulated widths proved impossible to locate using MOLSCAT and must be much
		narrower than 1~nG. Their tabulated positions are those of the threshold
		crossings identified by FIELD. Because of mixing between bound states, the
		mapping between resonances on the G2017 and F2013 potentials is uncertain for
		some resonances below 200~G.} \label{tab:predict40}
	
	\begin{tabular}{lr}
		\begin{tabular}{rcrccrrcrr}
			\hline\hline
			$B_{\rm res}$ (G) &\qquad& $B_{\rm res}$ (G)&\qquad& $\Delta$ (G) & \multispan3\hfill $a_{\rm bg}$ ($a_0$) \hfill& $L$ & $M_F$\\
			G2017&& F2013 && G2017 && G2017 \hfill &&\qquad\\
			potentials && potentials &&potentials&&potentials&&\qquad\\
			\hline
			31.002  &&  27.103  &&                      && $-66.9$  &&  2  &  +1/2 \\
			32.976  &&  29.035  &&                      && $-66.9$  &&  2  &  -1/2 \\
			35.308  &&  31.293  &&                      && $-66.9$  &&  2  &  -3/2 \\
			35.775  &&  31.611  &&                      && $-66.9$  &&  2  &  +1/2 \\
			38.048  &&  33.935  &&                      && $-66.9$  &&  2  &  -5/2 \\
			38.760  &&  34.382  &&                      && $-66.9$  &&  2  &  -1/2 \\
			41.263  &&  37.034  &&                      && $-66.9$  &&  2  &  -7/2 \\
			42.324  &&  37.679  &&                      && $-66.9$  &&  2  &  -3/2 \\
			42.976  &&  38.160  &&                      && $-66.9$  &&  2  &  +1/2 \\
			46.592  &&  41.636  && $-2\times10^{-17}$   && $-66.9$  &&  2  &  -5/2 \\
			47.660  &&  42.366  && $-2\times10^{-17}$   && $-66.9$  &&  2  &  -1/2 \\
			51.748  &&  46.435  &&                      && $-66.9$  &&  2  &  -7/2 \\
			53.410  &&  47.546  && $-3\times10^{-14}$   && $-66.9$  &&  2  &  -3/2 \\
			54.539  &&  48.410  &&                      && $-66.9$  &&  2  &  +1/2 \\
			60.587  &&  54.030  && $-5\times10^{-14}$   && $-66.9$  &&  2  &  -5/2 \\
			62.447  &&  57.590  && $-8\times10^{-18}$   && $-66.9$  &&  0  &  -3/2 \\
			62.537  &&  55.439  && $-3\times10^{-14}$   && $-66.9$  &&  2  &  -1/2 \\
			69.698  &&  62.295  && $-3\times10^{-14}$   && $-66.9$  &&  2  &  -7/2 \\
			72.924  &&  64.609  && $-3\times10^{-11}$   && $-66.9$  &&  2  &  -3/2 \\
			75.492  &&  66.582  && $-2\times10^{-15}$   && $-66.9$  &&  2  &  +1/2 \\
			75.524  &&  69.853  &&                      && $-66.9$  &&  0  &  -3/2 \\
			86.780  &&  76.911  && $-9\times10^{-11}$   && $-66.9$  &&  2  &  -5/2 \\
			91.630  &&  80.518  && $-8\times10^{-11}$   && $-66.9$  &&  2  &  -1/2 \\
			96.335  &&  89.010  && $-6\times10^{-14}$   && $-66.9$  &&  0  &  -3/2 \\
			105.702  &&  93.892  && $-1.4\times10^{-9}$  && $-66.9$  &&  2  &  -7/2 \\
			114.882  && 100.745  && $-2\times10^{-8}$    && $-66.8$  &&  2  &  -3/2 \\
			124.087  && 107.659  && $-2\times10^{-12}$   && $-66.8$  &&  2  &  +1/2 \\
			133.522  && 122.768  && $-6\times10^{-10}$   && $-66.8$  &&  0  &  -3/2 \\
			149.344  && 131.536  && $-2\times10^{-7}$    && $-66.8$  &&  2  &  -5/2 \\
			165.689  && 146.133  && $-8\times10^{-10}$   && $-66.7$  &&  2  &  -7/2 \\
			169.871  && 149.152  && $-6\times10^{-8}$    && $-66.7$  &&  2  &  -7/2 \\
			171.606  && 148.042  && $-1.3\times10^{-6}$  && $-66.7$  &&  2  &  -1/2 \\
			173.035  && 152.632  && $-1.0\times10^{-6}$  && $-66.7$  &&  2  &  -7/2 \\
			176.113  && 156.289  && $-9\times10^{-7}$    && $-66.7$  &&  2  &  -7/2 \\
			176.689  && 157.129  && $-1.4\times10^{-10}$ && $-66.7$  &&  2  &  -5/2 \\
			181.215  && 160.074  && $-7\times10^{-10}$   && $-66.6$  &&  2  &  -5/2 \\
			185.920  && 163.642  && $-4\times10^{-8}$    && $-66.6$  &&  2  &  -5/2 \\
			190.232  && 167.934  && $-4\times10^{-7}$    && $-66.6$  &&  2  &  -5/2 \\
			190.509  && 170.875  && $-3\times10^{-10}$   && $-66.6$  &&  2  &  -3/2 \\
			194.198  && 172.611  && $-3\times10^{-7}$    && $-66.6$  &&  2  &  -5/2 \\
			195.011  && 173.418  && $-1.3\times10^{-8}$  && $-66.6$  &&  2  &  -3/2 \\
			199.611  && 176.821  && $-4\times10^{-7}$    && $-66.5$  &&  2  &  -3/2 \\
			\hline\hline
		\end{tabular}\qquad&\qquad
		\begin{tabular}{rcrccrrcrr}
			\hline\hline
			$B_{\rm res}$ (G) &\qquad& $B_{\rm res}$ (G)&\qquad& $\Delta$ (G) & \multispan3\hfill $a_{\rm bg}$ ($a_0$) \hfill& $L$ & $M_F$\\
			G2017&& F2013 && G2017 && G2017 \hfill &&\qquad\\
			potentials && potentials &&potentials&&potentials&&\qquad\\
			\hline
			203.797  && 181.086  && $-5\times10^{-6}$    && $-66.5$  &&  2  &  -3/2 \\
			208.400  && 185.944  && $-9\times10^{-6}$    && $-66.4$  &&  2  &  -3/2 \\
			208.414  && 188.721  && $-1.2\times10^{-10}$ && $-66.4$  &&  2  &  -1/2 \\
			208.624  && 180.762  && $-5\times10^{-5}$    && $-66.4$  &&  2  &  -7/2 \\
			212.577  && 190.265  && $-9\times10^{-9}$    && $-66.2$  &&  2  &  -1/2 \\
			212.873  && 192.186  && $-9\times10^{-4}$    && $-66.4$  &&  2  &  -3/2 \\
			215.626  && 196.709  && $-2\times10^{-5}$    && $-66.4$  &&  0  &  -3/2 \\
			217.044  && 192.995  && $-7\times10^{-7}$    && $-66.4$  &&  2  &  -1/2 \\
			221.848  && 196.981  && $-6\times10^{-6}$    && $-66.3$  &&  2  &  -1/2 \\
			226.715  && 202.190  && $-8\times10^{-5}$    && $-66.2$  &&  2  &  -1/2 \\
			232.248  && 211.942  && $-5\times10^{-9}$    && $-65.8$  &&  2  &  +1/2 \\
			232.589  && 208.400  && $-7\times10^{-4}$    && $-66.0$  &&  2  &  -1/2 \\
			236.460  && 213.627  && $-5\times10^{-9}$    && $-64.9$  &&  2  &  +1/2 \\
			237.183  && 215.961  && $-0.013$             && $-66.0$  &&  2  &  -1/2 \\
			239.858  && 214.741  && $-5\times10^{-9}$    && $-66.3$  &&  2  &  +1/2 \\
			243.371  && 216.624  && $-4\times10^{-8}$    && $-66.0$  &&  2  &  +1/2 \\
			247.638  && 221.339  && $-2\times10^{-7}$    && $-65.8$  &&  2  &  +1/2 \\
			250.777  && 230.247  && $-2\times10^{-8}$    && $-65.6$  &&  0  &  -3/2 \\
			253.115  && 227.532  && $-2\times10^{-6}$    && $-65.5$  &&  2  &  +1/2 \\
			256.464  && 234.147  && $-1.1\times10^{-6}$  && $-65.3$  &&  0  &  -3/2 \\
			257.695  && 235.275  && $-2\times10^{-5}$    && $-65.2$  &&  2  &  +1/2 \\
			262.890  && 239.549  && $-4\times10^{-5}$    && $-64.7$  &&  0  &  -3/2 \\
			265.979  && 224.767  && $-9\times10^{-5}$    && $-64.3$  &&  2  &  -3/2 \\
			270.122  && 246.441  && $-1.4\times10^{-3}$  && $-63.4$  &&  0  &  -3/2 \\
			277.215  && 254.519  && $-0.021$             && $-60.7$  &&  0  &  -3/2 \\
			285.964  && 264.340  && $-0.86          $    && $-67.5$  &&  0  &  -3/2 \\
			411.372  && 318.650  && $-2\times10^{-9}$    && $-67.8$  &&  2  &  +1/2 \\
			441.876  && 379.663  && $-0.002$             && $-67.6$  &&  2  &  -5/2 \\
			489.071  && 466.458  &&                      && $-67.6$  &&  2  &  +1/2 \\
			531.217  && 470.254  && $-0.055$             && $-67.7$  &&  0  &  -3/2 \\
			555.460  && 531.469  &&                      && $-67.8$  &&  2  &  -1/2 \\
			613.799  && 588.830  &&                      && $-67.7$  &&  2  &  -3/2 \\
			661.323  && 627.835  &&                      && $-67.7$  &&  2  &  +1/2 \\
			667.781  && 642.012  &&                      && $-67.7$  &&  2  &  -5/2 \\
			704.214  && 677.444  && $-3\times10^{-18}$   && $-67.7$  &&  0  &  -3/2 \\
			718.940  && 692.463  &&                      && $-67.7$  &&  2  &  -7/2 \\
			741.968  && 706.122  &&                      && $-67.7$  &&  2  &  -1/2 \\
			817.559  && 779.664  &&                      && $-67.7$  &&  2  &  -3/2 \\
			890.432  && 850.612  &&                      && $-67.7$  &&  2  &  -5/2 \\
			944.619  && 902.844  && $-7\times10^{-17}$   && $-67.7$  &&  0  &  -3/2 \\
			949.117  && 892.623  &&                      && $-67.7$  &&  2  &  +1/2 \\
			961.839  && 920.113  &&                      && $-67.7$  &&  2  &  -7/2 \\
			\hline\hline
		\end{tabular}
	\end{tabular}
\end{table*}

\end{document}